# Unification of computer reality

**Sergey I. Kruzhilov**

PhD, associated professor

National Research University "Moscow Power Engineering Institute",
ksi1048@gmail.com

**Abstract.** The work attempts to unify the conceptual model of the user's virtual computer environment, with the aim of combining the local environments of operating systems and the global Internet environment into a single virtual environment built on general principles. To solve this problem, it is proposed to unify the conceptual basis of these environments. The existing conceptual basis of operating systems, built on the "desktop" metaphor, contains redundant concepts associated with computer architecture. The use of the spatial conceptual basis "object - place" with the concepts of "domain", "site", and "data object" allows to completely virtualize the user environment, separating it from the hardware concepts.

The virtual concept "domain" is becoming a universal way of structuring the user's space. The introduction of this concept to describe the environments of operating systems provides at the mental level the integration of the structures of the local and global space. The use of the concept of "personal domain" will allow replacing the concept of "personal computer" in the mind of the user. Concept "partition" of the personal domain will replace concepts "desktop" and external "memory device".

The virtual concept of "site" as an environment for activities and data storage will allow abandoning such concepts as "application" (program), or "memory device". The site in the mind of the user is a virtual environment that includes both places for storing data objects and places for working with them. Integration of programs and data within the site can be carried out using portals. The software component is replaced by the concept of a "tool", and the concept of "data" is no longer associated with memory devices. Data becomes virtual objects stored in virtual data sites ("storages"). The data structure is displayed on the desktops of these sites. For the search and logistics of data, special tools are used. The introduction of the "storage" concept allows not to separate data from the means of their management. The introduction of the concept "site" into the structure of operating systems environments and the concept of "data site" into the structure of the global network integrates the structure of the global and local space in the user's mind.

The introduction of the concept of "portal" as a means of integrating information necessary for interaction, allows ensuring the methodological homogeneity of the user's work in a single virtual environment. From the user's point of view, the way of interacting with elements of the environment will depend only on the conceptual type of these elements and will not depend on where they are located. The portal is also a means of integrating application sites with data sites. With the help of portals that hide the features of the interaction process from the user, the user's methods of work in the local environment of his personal domain become similar to the methods of work in the global network.

The similarity of the structures of global domains, personal domains, and data sites provides a similarity in the methods of user interaction both with sites and with data at all levels.

## Introduction

In this work, an attempt is made to unify the conceptual models of user's virtual computer environments in order to combine the local environments of operating systems and the global Internet environment into a unified virtual environment built on common principles. The used methodology is systems theory. Certain elements and principles of organizing this environment are already being used in practice. The task is to bring them together into a unified system.

## Mixed reality

Mixed reality is a natural human environment. The basic principle of human existence is "goal - means". A person constantly solves the tasks of survival in the environment, using the resources of physical reality as a means. In the process of life, he sets and solves various tasks. The reality in which a person exists is mixed (Figure 1).

On the one hand, a person as a physical entity in the process of his activity interacts with other objects of physical reality, and on the other hand, this interaction is impossible without the use of symbolic objects of the virtual reality of his consciousness. Objects of physical and virtual reality are interconnected. A person perceives information about physical objects in the form of virtual images, evaluates this information, which makes it possible to adequately respond when interacting with environmental objects.

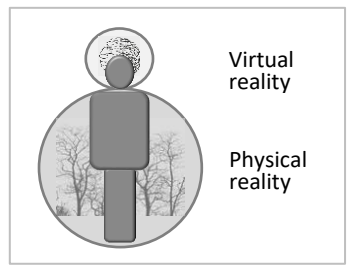

Figure 1. Mixed reality

A generalized model of the process of interaction of these objects in solving various problems in the environment is shown in Figure 2. In the process of sensation with the help of the senses, a person perceives the properties of objects in a signed form, data on the state of the environment. Disparate sensations are connected by the human cognitive apparatus into a single perceptual image, which is the essence of the concept of "information".

The basic principle of human association is "contour - background". A person automatically selects objects, entities, bounded by a contour from the general background of the environment. The contour allows you to recognize objects at a great distance when individual details of the image are not yet visible. The contour principle of perception underlies discrete models for describing the world by a person. Due to the principles of the cognitive apparatus, a person's perception of the world is object-oriented. The main categories for describing physical reality are the categories "object" and "space".

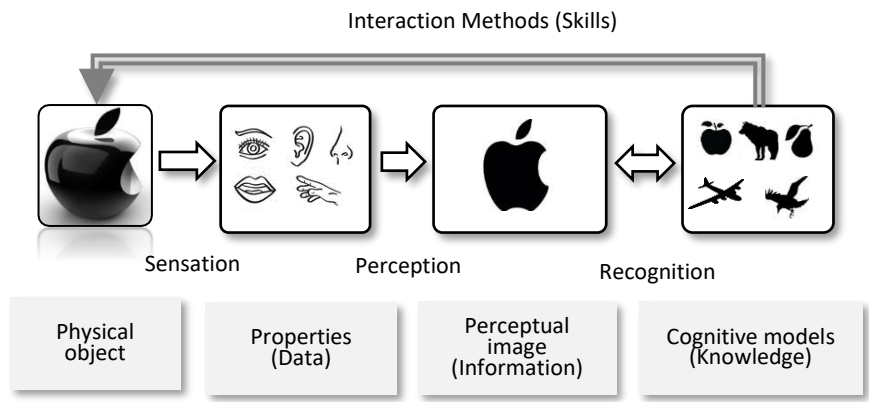

Figure 2. A generalized model of the process of interaction

The opposition "object - space" is conditional. Whether to count something as an object or space depends on the point of view of the observer. Since any object is a part of space bounded by a contour, its perception is dual. Under certain conditions, an object can be perceived as space, and space can be perceived as an object. When a person's attention is transferred from the contour of the object to his internal structure, the object turns into a limited space (place). As an example, we can cite the dual perception of a monitor: on the one hand, it is a limited physical object, and on the other hand, it is a space for displaying symbolic objects of the virtual environment on the screen.

The principle of association "part - whole" allows you to combine a group of objects into a single image. For example, integrate details of the face and clothing into a single image of a person. The association of a group of objects into a single whole gives rise to the concept of "structure". Our cognitive apparatus automatically provides us with a structured perception of the world. Not only objects are structured, but also the space of the environment, perceived as a structure of separate subspaces. The perception of a complex structure as a single object is the essence of the "system" concept (literally a complex object). The way data is associated is not straightforward and depends on the point of view of the observer. An illustrative example is the well-known Necker cube with two modes of perception (Figure 3).

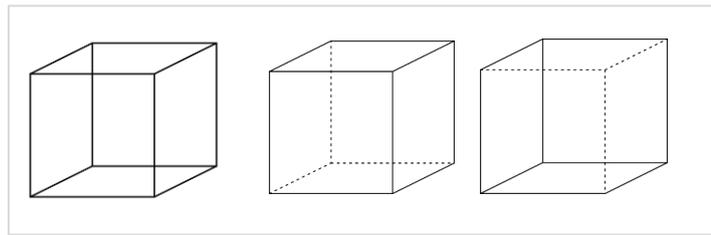

Figure 3. Necker cube

An adequate response to perceived information is impossible without recognition of perceptual images. Recognition is carried out by comparing the information image with cognitive models stored in long-term memory. These models are formed in the process of training human memory cells during multiple interactions with a group of similar physical objects. In parallel, in the learning process, methods of interaction with objects of this type are formed.

Cognitive or mental models determine the semantic of perceived information and are the basis of our knowledge. A lot of methods of interaction forms the basis of our skills. Interaction methods are directly related to the concept of an object's "function", the way a person uses an object. An object's function is also relative. The same object can be used in different ways depending on the tasks of the person. Without mental models and methods, human interaction with physical objects is impossible.

Human perception is abstract. Human perceived properties, information images and mental models are generalized. For example, "red" is not light of a specific wavelength, but of a whole range of waves. An informational image is not a static photograph, but a dynamic image that integrates views of an object from different angles. Any model is a collective image corresponding to a class of objects. The methods of interaction are also generalized. Our mental world is a world of abstractions. Abstraction reduces memory load. Abstraction underlies the human ability to classify objects.

A person has the ability to internally visualize his virtual objects in representation and imagination. These visual images form the basis of our consciousness. Imaginary virtualization enables a person to create "chimeras" by combining familiar images in the mind into new systems. Virtual combinatorics allows us to establish causal relationships and create new artificial systems. The same combinatorics at the level of methods underlies the planning and programming of human activities in solving problems.

Virtual reality objects of human consciousness are closely related to objects of physical reality. Figurative mental models determine the meaning of physical objects for us. Various figurative objects form the basis of human conscious activity. But consciousness is just the tip of the cognitive iceberg. Most cognitive activity occurs unconsciously. Nevertheless, it is conscious activity that underlies the creative activity of a person and the transfer of social experience. It is a conscious activity that allows the user to navigate through the iconic space of a virtual computing environment. Although individual actions can be performed by him at a subconscious level (skills).

The cognitive capabilities of human consciousness are limited. The complexity of the structures of mental models used by a person, as a rule, does not exceed 7 elements (for most people, no more than 3-5 elements). Exceeding the limit of complexity leads to the transformation of a conscious virtual structure into an unstructured background. The complexity of the structures of perceptual information images is also limited. The ultimate complexity of such structures is no more than 20 elements (for most, no more than 12-16 elements). An increase in the complexity of cognitive structures is possible with a hierarchical disclosure of complexity from the shift of the focus of attention from the structure of the whole to the structure of constituent elements. Hierarchy significantly increases the complexity of cognitive images. The depth of the hierarchy is limited: for mental models - approximately 2-3 levels, and for perceptual signs - up to 7 levels. Hyperstructural perception is a consequence of the limited cognitive capabilities of a person. Consideration of cognitive limitations is mandatory when designing interfaces for software systems.

## Sign systems

Virtual space is a person's inner world, inaccessible to other people. When transferring experience to exchange accumulated information, a person uses artificial sign systems. A sign is a perceived object that carries information about another object (value object). The general relationship of these objects is shown in Figure 4. Like physical objects, signs are associated with mental models in the mind of a person. A cognitive model of a sign, like a cognitive model of a physical object, is the semantics of a sign. Signs can denote both objects of physical and virtual reality. Sign objects are augmented reality of a person's mixed reality that occupies an intermediate position between physical and virtual reality.

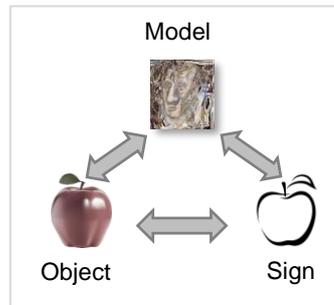

Figure 4. Sign structure

The widespread use and development of sign systems is a specific path of human evolution. The most important stages of this evolution are speech, writing, painting, drawing, photography, cinematograph, radio, TV and, finally, interactive sign computer environments.

The use of sign systems to describe reality is a transition from mental models to perceptual models, with the corresponding possibility of increasing the complexity of perceived structures from 7 to 20 elements at each level. Thus, the use of drawings in the design has significantly increased the complexity of the designed systems.

Since mental models are abstract in nature, the sign usually denotes not one object, but a class of objects. The triple "sign - virtual model - class of objects" is the basis of the logical category "concept". Concepts are the main tool for describing reality. The conceptual basis of a subject area is a set of concepts necessary to describe this area. Categories are the most general concepts. Due to the limited cognitive capabilities of a person, the conceptual basis used by him is always limited. With the complication of the conceptual basis, the effectiveness of human activity is lost. For example, an increase in the number of types of tools or objects in a software system complicates access to them and leads to a loss of user efficiency. The simpler the basis, the faster the adaptation to the environment, the higher the efficiency of human activity in the environment.

A person has only one cognitive apparatus, so interaction with iconic objects of augmented reality is carried out in the same way as with other objects of mixed reality. Accordingly, each signed object is usually associated with its own conceptual type with an abstract mental model and generalized methods of interaction. Signs can also be associated with their specific properties and structure. Images of iconic objects are usually primitive, which leads to the problem of distinguishing between objects of the same type by the user.

## Conceptual basis of physical reality

The conceptual basis of the physical reality of a person "end-means" is determined by the general principle of his existence in the environment. To survive, a person sets goals and seeks the means of their realization. In accordance with this principle, the main categories are the category "task", as a realizable goal, and the category "object - place", to describe the means of solving the problem. The main object for a person is another "person". In addition, for the implementation of the tasks, a person uses "objects" of the environment as means, including tools, objects of influence on other objects. Human activity is usually carried out in specialized "places" (sites) containing all the necessary items and tools.

These are places for work, study, recreation, and entertainment: homes, workshops, warehouses, schools, libraries, clubs, stadiums, theaters, etc. People interact at communication platforms. The implementation of complex tasks involves the solution of several subtasks and the presence for each subtask of a specialized workplace with its own toolkit. For storage of objects, special storage areas are used. The solution of complex tasks also presupposes the need for a special place for managing subtasks. The effectiveness of the task implementation essentially depends on the way of organizing the structure of the workspace.

Problem-solving sites are in specific locations, usually owned by individuals. Access to these sites is most often restricted. Examples of such places can be private households, territories of enterprises and settlements, regions, or countries. Zoning is a standard way to improve the efficiency of finding the right sites. Territory selection reduces the search area. Zoning is used not only when searching for sites, but also when searching for the desired objects in storages or tools in toolboxes. Vehicles may be required to move people and objects from site to site.

Thus, the conceptual basis for describing human activity in physical reality is described by the categories "task", "object" and " place". The category "object" includes such concepts as" person"," object "as a means, and "tool". The "place" category includes such concepts as "site", "workplace", "desktop", and "domain" as a controlled territory.

## The heterogeneity of augmented reality

The current structure of computer augmented reality is not homogeneous. It consists of different environments with different conceptual bases. The first group of environments is the spaces of various operating systems. Another environment is the global space of the Internet.

The conceptual basis of the virtual environments of modern computer operating systems is formed in accordance with the well-established metaphor of the "desktop". This basis was formed in a specific scientific environment of computer specialists. Therefore, it is closely related to such concepts of computer architecture as "computer", memory "device", "RAM", "program" (application). To describe the data in this framework, we use "office" concepts related to the structure of the standard workplace of a researcher: "file", "folder", "desktop", which is more a repository of files and folders than a desktop with tool sets.

The structure of the global WWW-environment is built on a conceptual basis that does not use the concepts of computer architecture: "domain", "site", "hypertext" (book hypertext model of a scientific monograph or article). These concepts are closer to the familiar basis of most people.

Additional computer reality takes an increasing place in people's lives every year. The conceptual basis currently used when working in a computer environment is redundant and overloaded with concepts of computer architecture that are unnecessary for the work of most users. Virtualization of computer reality and its separation from hardware support is a modern trend in the development of the user environment. Combining various spaces into a unified homogeneous space, with a common conceptual basis "task, object, place", will simplify the structure of the environment and increase the efficiency of the user's activity.

## The principles of unification of augmented computer reality

Augmented computer reality is not isolated. It is closely related to both the physical and mental virtual reality of a person. Signed objects of computer reality can be transformed into physical objects using various devices (3D printers, digital machine-tools). Conversely, using various scanners, physical objects can be converted into virtual objects. With the help of physical objects (mouse, keyboard, hand ...), the user can act on virtual objects on the monitor screen, which in turn allows him to control real physical systems and processes.

Due to the close interconnection of the physical and sign media, it is desirable to ensure the similarity of their conceptual bases. Since physical reality is primary for a person, the conceptual basis of computer reality should be approximated to physical reality, and not vice versa. Augmented computer reality should be built on the same conceptual basis as the physical one.

Nowadays, WWW sites are acquiring the features of full-fledged applications. The user's data, which was previously local, is now stored not on computer memory devices only, but in the cloud environment of the global network. Strengthening the relationship between the local environments of operating systems and the global Internet environment requires mutual unification of their conceptual basis.

Since the objects of computer reality are sign objects, the unification of user augmented reality should be carried out not only at the conceptual but also at the sign level. Both the global WWW environment and the local operating system environments should look the same from the user's point of view, as part of a unified user environment. Accordingly, the methods of interaction with sign objects in different parts of the computer reality should also be similar. These methods should only depend on the conceptual type of the object and should not depend on its location.

# Unification of virtual computer reality at the concept level

Conceptual unification of the user's environment can be carried out within the framework of the standard "goal-means" paradigm using the usual conceptual basis of human physical reality: "tasks" as a process of achieving a goal, "persons" as a means for jointly solving tasks, "objects" as a means for solving a problem, "tools" as a means of influencing objects, "sites as a place for solving problems, and "domains" as a location for sites.

With this approach, the global computer network is mentally perceived by the user as a common sign space described by a common spatial object-oriented conceptual basis. The concepts of the computer architecture of this network fade into the background. Virtualization of the user environment with the abandonment concepts of the computer architecture is a modern trend in the development of the user interface. Today, the concepts of "computer" and "device" are increasingly virtualized. User resources can be located on different computers, and the same computer can contain the personal environments of different users. In such conditions, knowledge of computer architecture is not necessary to solve the user's problems in the virtual environment of his additional reality.

Consider the structure of the main categories of the user's the conceptual basis.

## Domain

From the user's point of view, the structure of a unified space (see Figure 5) consists of "domains", managed subspaces belonging to different users. Domains contain "sites", places with means for activities. Domain access is controlled by its owners. At the top level, a unified space consists of a group of global domains. Corporate domains and personal user domains are located within global domains. Personal domains of users are their personal environments, implemented both by means of their personal computers and by network means (cloud resources). When a computer is connected to the global network, personal domains become part of the global space. A personal domain is a virtual concept. The user can have access to any available resources of the single space. The user experience with environment resources is the same regardless of where they are located.

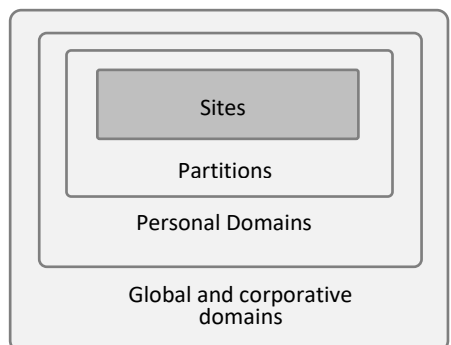

Figure 5. Domain structure

To improve the efficiency of access to sites, a personal domain can be divided into virtual partitions. Domains and partitions are a spatial way of implementing a classifier when the class of objects is determined by their location (association of objects by contiguity). Selecting a domain or partition narrows the search scope. Partitions can be located not only on the permanent media of a computer but also on mounted ones. The connection of an external memory device is perceived by the user as the appearance of another partition in the structure of his personal domain. User partitions can be located in the "cloud".

The concept of "personal domain" makes it possible to make the concept of "personal computer" optional. The personal domain environment should not be hardcoded to a specific computer. If necessary, it can be transferred from one computer to another in the same way as today the user's environment is transferred from one smartphone to another.

The user's personal domain is managed through the system site. Using the tools of this site, you can create and complete user tasks, provide access to the resources of a unified environment, and also change the structure of a personal domain. In terms of its functions, the system site is in many ways similar to a WEB browser.

## Site

Sites are places to accomplish user tasks with sets of tools necessary for this. The site is not a program, but a virtual spatial structure, similar to the concept of an "application window". The concept of "site" is closely related to the concept of "task", since in a virtual computer environment, any site is a means of solving problems of a certain class, and the created task is always associated with the corresponding site.

The concept of a task is broader than the concept of a site. A complex task may require solving several subtasks, using the tools of other sites and specialized workplaces. Therefore, to solve complex problems, the user must have task management tools at each site. The solution to any problem involves the use of appropriate data objects as the necessary means. That is, each site is usually associated with its own local database with data management tools. The solution of specific subtasks is carried out in specialized workplaces with tools and desktops included in the site structure. Typically, task and data management are the prerogative of the system that controls access to personal domain resources.

Since the places for managing tasks and data are used to solve almost all tasks, from the user's point of view, they can be considered as workplaces of almost every site. The inclusion of these places in the list of workplaces of any site makes it possible to consider each site as a full-fledged local environment for solving the user's problem. The user does not need to know the solution of which subtasks are outsourced to the system, and which tasks are solved by the means of the site itself.

The general structure of the site's workplaces is shown in Figure 6. Each workplace consists of a desktop for displaying the data structure and a toolkit, placed on the toolbars.

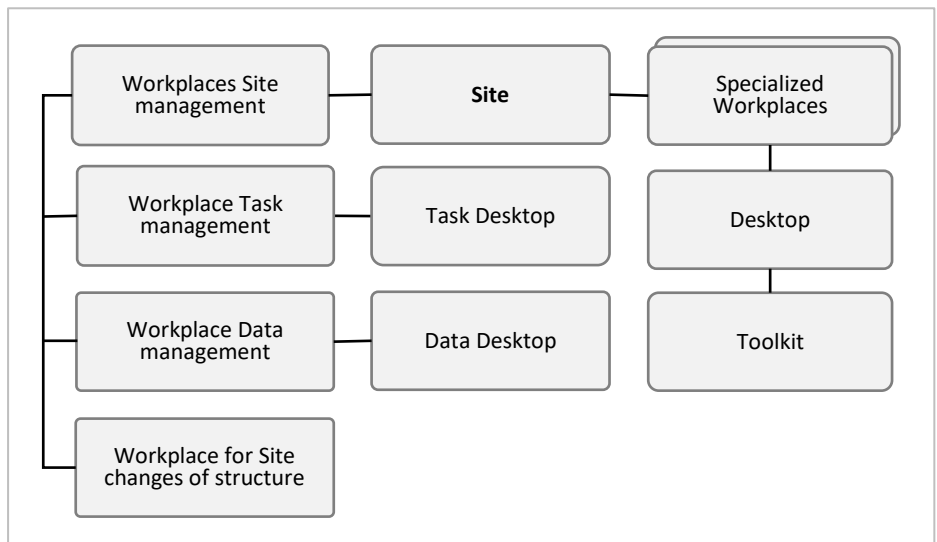

Figure 6. Site structure

The site is a universal spatial way of presenting both programs and data. The software component is represented by the functions of a set of virtual instruments implemented by software agents hidden from the user, and the data is represented by a virtual spatial structure on the data management desktop, with memory devices hidden from the user for storing them. With this approach, the user may not use the concept of "program (application)", which is closely related to the concepts of processor and random-access memory, as well as the concept of "memory device". Both programs and data are replaced by virtual spatial structures that are not associated with specific physical devices. A partial analog of the concept of "site" is the concept of " window".

By analogy with physical reality, user data objects are stored in data sites (storages). The data site is a virtual platform with a desktop for displaying the user's data structure and a set of data search and logistics tools. During work, user data objects are not moved from the device to RAM, but from the data site to the desktop and, at the end of the work, are returned back to the data site. From the data site, data objects can be moved to other data sites. Data sites can exist as separate units or be associated parts of other sites in the form of data management workplaces. The data site presupposes both its own data management workplace and a place for job management to implement the logistic functions of data movement.

Unlike the desktop metaphor with a single user file system, data storage is a means of localizing user data with effective access to data objects and a means of protecting data from unauthorized access. This principle of working with objects is characteristic of both physical reality and the WWW-environment.

The structure of the data site resembles a container terminal and is described by the concepts of "data storage", "section" of the storage, "container" and "object" of data (Figure 7). In essence, the storage structure is similar to the structure of a personal domain. At the top level, the storage consists of a set of virtual partitions for dividing the storage into separate zones. Like partitions of a personal domain, these zones are a means of speeding up access to objects during manual searches. "Container" is a means of grouping data with the ability to move them together. The role of the container in the data structure is similar to the role of the folder concept in the file system. Containers can be nested within other containers, but the nesting level should be limited. User data objects are located in partitions and containers.

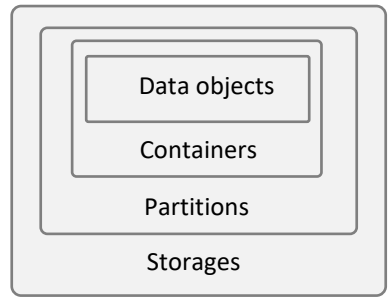

Figure 7. Data structure

The storage associated with the site can store not only data objects but also the current site settings, as well as work logs that provide the ability to recover the results of work in case of system failures or user errors. It can also store a local trash container for deleted objects, as well as a local clipboard.

The storage associated with the application site can serve as a means of restricting access to user data during remote work. The external agent has the ability to access data only within the associated storage. Access control to the storage of the user's objects is always carried out by means of the system of his personal domain. Access to data when working with remote storage is always implemented by the data management system of its personal domain, capable of interacting with agents of data sites in other personal domains.

### Object

The user's objects are people, tools, and data objects. People are objects for joint problem-solving. Coordination of human activities is carried out using communication means. Tools are objects that have a function implemented by a hidden software agent.

Data objects are tools for solving user problems. They can be used when working on sites or created on specialized sites. Data objects are classified by types: documents, drawings, photographs, programs, etc. An object type is associated with its conceptual model and interaction methods. Both the model and the methods are stored in the user's mental memory. Virtual data objects are also characterized by their properties and structure. Properties and structure are stored in data structures associated with the sign of the object. Objects are stored in storage, and their structures are displayed on the desktops of the respective sites.

In general, a data object is an aggregate of several interconnected objects stored in one package. Examples of such aggregates are movies in the mkv-video, which include a video track, audio tracks, and various subtitles, or mp4 or jpeg media containers. Any program as a set of interconnected program modules is also an aggregate. Concept "file" refers to the packaging rather than the object it contains.

### The principles of organizing the site structure

The principle of organizing work on the site significantly affects the efficiency of solving problems by the user. An important criterion for work efficiency is the user's time and effort to find the right tools. Each such search is a distraction from the solution to the problem. The user can access the tools directly using commands. The command line is still one of the most common ways to activate functions. The disadvantage of this method is the limited number of commands due to the limited cognitive capabilities of the user. An increase in the number of commands leads to an increase in the number of errors when typing them.

With a large number of tools, a classification principle is used to access the tools. The main classification principles used in practice are technological and genus-species. In accordance with the technological principle, tools are classified according to the technological stages of the work. Each of these steps is performed at a separate workplace with its own desktop and a limited toolkit. The genus-species classification principle involves dividing a set of instruments into non-overlapping subsets in accordance with their common properties ("fruits to fruits", and "vegetables to vegetables"). The standard way to do this is through a menu. The problem with the genus-species approach is that it is often difficult to offer a user-understandable principle of classifying tools. In real production, the technological approach dominates, and in the virtual computer environment, the genus-species one. In practice, various combinations of these approaches are possible.

The difference between the described approaches can be illustrated by the example of the organization of the MS Word text editor environment, where the tools are divided according to the classification principle into non-overlapping classes according to a very specific criterion: File, Home, Menu, Insert, Design, Page Layout, References, Mailings, Review, View. Selecting a class changes the set of tools on the ribbon (analogous to the toolbar). When accessing the desired tool, you must first understand what class it belongs to. Finding many instruments is not a trivial task in this classifier. For example, which of these sections should you look for the Hyphenation tool?

Access to tools using a technological approach can be more effective. With this approach, it is more logical to divide the process of creating a document into stages in accordance with the technology of constructing the structure of the document. The number of conceptual types of any document is limited: text, pictures, tables, formulas, and links. Each of these elements could be created and modified on a separate desktop with a specific limited set of tools. To create a document you need a restricted set of workplaces: "Document" - to work with the document as a whole (opening, saving, printing ...), "Text" - to enter and format text, "Table" - to create tables, "Figure "- to insert and edit pictures, "Formula "- to create formulas and" Reference "- to insert links. In this case, each executed subtask limits the set of tools used. If the number of instruments is still large, then they can be divided into groups. Understanding the technology of document creation, the user understands where the necessary tools can be located.

## Unification of virtual computer reality at the sign level

From the user's point of view, virtual computer space is a set of limited symbolic objects displayed in the space of the monitor screen. The user interacts with these objects when solving his problems, just as he interacts with real objects in physical space. Since the structure of a unified environment of virtual reality is implemented within the framework of the general paradigm "goal - means", and means within the framework of the paradigm "object - place", then in this structure it is necessary to provide both signs for designating tasks and signs for designating objects and places.

As in the physical environment, the user, by the appearance of the sign, must determine its conceptual type, which will give him the opportunity to understand what it is and how to interact with it. A sign object, like a physical one, is characterized by a set of properties. Since at best the conceptual type of an object is determined by its appearance, the properties of the sign have to be stored in a special data structure associated with the sign and given to the user at his request. A sign can be associated with data, carrying information about its internal structure. The structures of such objects are visualized on the desktops of specialized workplaces. The sign can have its own function, which is implemented by the associated software agent.

To distinguish between signs of the same type, unique identifiers are used. The identifier is one of the properties of the object, usually hidden from the user. The identifier is assigned by the system automatically. An additional property used by the user to distinguish between signs in a bounded context is the personal name of the sign, given by the user and visible to him. The name is not required to be unique. Various tools are used to access the various components of the iconic object. The general structure of the sign is shown in Figure 8.

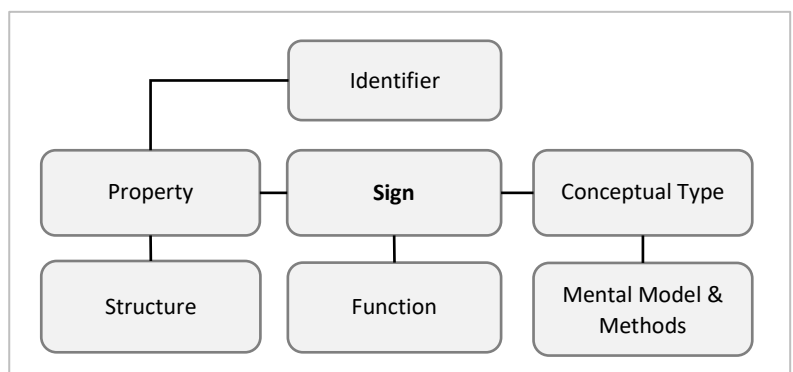

Figure 8. General structure of the sign

Signs represent individuals as well as tools, tools, and data objects. Portal signs are used to indicate tasks and locations.

Personality signs are associated with information about the relevant people and how to communicate with them. This information is contained in calling cards of individuals stored in address books or user profiles. To organize interaction, people can exchange business cards.

Data object signs are symbols associated with objects that have a structure. Examples of such signs are signs of photos, drawings, videos, documents, etc. The structure of such objects is visualized and changed on the desktops of special workplaces. Data object signs are denoted user's objects in data storages.

Tool signs are signs for objects associated with functions implemented by hidden software agents. The sign function is activated by acting on the sign of the tool. Toolbars with tool signs are an important part of the site structure.

Portal signs are signs for tasks and spaces. Portals not only designate a place in the virtual space but at the same time are a "vehicle" for moving the user to these places. Activation of the portal leads to a change in the sign structure on the monitor screen, which is perceived by the user as moving to another part of the space. Since in a computer environment the task is closely related to the site, the tasks are also indicated by the signs of the portals. When working in multitasking mode, switching to the desired task is the activation of the corresponding portal with the transition to the space of the site associated with the solution of the task.

Signs of portals indicate tasks, domains, sections, sites, and site jobs. In storages portals are used to denote partitions and containers. Portals can be used to denote parts of data objects. In the latter case, portals are a means of gradual disclosure of the structure of a complex object in the desktop space from the general structure to the structure of its details. That is, the portal is a means of forming hierarchical hyperstructures. From the user's point of view, the map of a unified virtual space is a hierarchical hyper-network of portals that designate domains and sites. Maps of sites have a similar hyper-network structure. The complexity of these maps should not exceed the cognitive capabilities of the person. Since the portal is the entry point to a certain space, an exit portal must be provided inside each space, returning the user to the original place. The exit can be carried out both with the help of the exit portal and using the "Exit" key on the keyboard or on the case (analog of the Escape key).

The transition to a new space is always a new task for the user, requiring orientation in a new environment and making a decision on further actions. This task may or may not be displayed by the portal on the tasks desktop. An analog can be a transition to a new page of the site in the browser, which can be displayed as a new task in a new tab, or as a new task in an old tab. But in any case, the task is necessarily recorded in the task log. A task journal is an important tool for organizing work. The chain of interrelated tasks with the transfer of parameters from task to task provides the possibility of decentralized work to achieve the set goal.

The concept of "portal" is an extension of the concept of "reference". The difference between a portal and a reference is that the portal is a means of associating programs with data and, as a rule, contains several references to the necessary resources. In the general case, the structure of the portal can include references to interface agents, software agents of sites, pointers to data sites necessary for operation, pointers to transmitted parameters, and communication agents. It is allowed to create portals with references to other portals. There is only one level of the second-level portal hierarchy. A reference to reference is the reference to the original portal.

Saving the context of work in the partitions of the associated storage makes it possible to ensure the reliability and stability of the user's work in case of possible system failures or errors in work. When working remotely, access to user data for external agents is limited to the context of the specified storage. Since the portal is associated with the context of work, the portals for working with the same site, but with different contexts will be different.

Portals are a means of realizing the methodological homogeneity of interaction with objects. Their use makes it possible to ensure both uniformity of access to the resources of the virtual space and the uniformity of methods of interaction with these resources. The uniformity of access means that you can get to any part of the space only through the appropriate portal. In addition, if you include in the structure of the portal information about the specifics of interaction with the agents, then it becomes a means of uniformity of interaction. The use of portals allows hiding from the user the specifics of organizing interaction with local and remote resources of the environment. With this approach, the methods of interaction with the resources of the environment will be determined only by the conceptual type of the corresponding sign object.

The user should not be burdened with the work of creating portals. Portals should be generated automatically at the user's request. For this, the system has all the necessary information. The maximum effort of the user is to come up with an appropriate personal name for the portal and, if necessary, add additional tags to the portal properties for the convenience of finding them in the personal virtual space. With the help of portals copied from other people's sites or sent to the user by the owners of resources, you can organize remote access to the resources of other users.

The implementation of methodological homogeneity with the possibility of local and remote access to the same resources presupposes the use both in the global WWW environment and in the local environment of operating systems of a network client-server approach to system interaction. Any user task is considered as a process of interaction between two agents: the user's interface client and the executive server (Figure 9). The task of the "client" agent is to organize user interaction with the iconic objects of the virtual environment. The task of "server" agents is to interpret user actions as commands. Clients are always part of the user's personal domain. Servers can be both parts of the user's personal domain and parts of a remote domain. With this approach, it is possible to provide both local and remote user access to resources by changing only the communication protocol and the data transfer route.

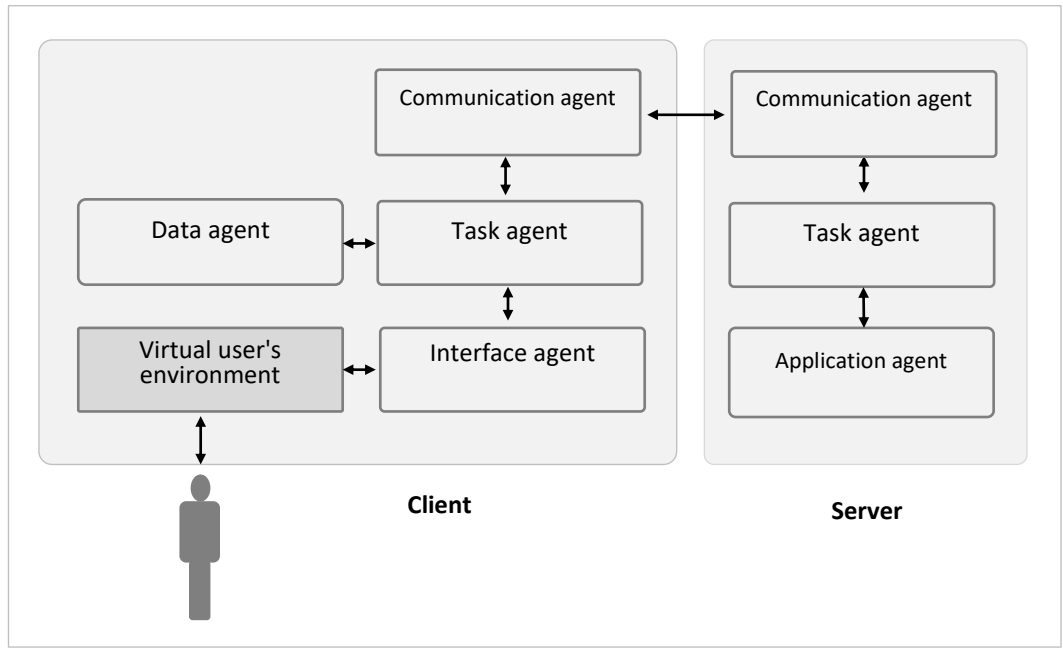

Figure 9. Client-server communication

Interface agents are system components of a user's personal domain. Since task and data management is a user's prerogative, task and data management agents are also system components of his personal domain. Workplace agents can be both parts of a user's personal domain server or part of a remote server. The interaction of the client and server agents is carried out by the communication agents, delivering packets with information from the client to the server and vice versa. In particular, such a communication network provides client-server interaction and within a personal domain.

Since the main types of threats come from executable applications, separating the client from the server within the personal domain will increase the security of the user's personal environment. The interface agent, like the task and data agents, is part of the secure client environment that the server application agents cannot change. Since the data is managed by the client data agent, the application agents can only access the resources to which the user has been granted permission.

Server agents send a vector graphic metafile to the interface agent, which is displayed by the client in the corresponding area on the monitor screen. When transferring data, various communication protocols can be used, GUI, HTTP, Blue-tooth, Wi-Fi, etc. Both the standard system interface vector graphics language (GUI) and the hypertext markup language (HTML) can be used as the description language. Since the interpreted description of a vector image is machine-independent, the interface description languages, in principle, can be any. For example, the GUI language of the local system can be used to interact with sites on the global network, and HTML to interact with the sites of user's personal domain. Each language assumes the presence of a corresponding interface agent that interprets the graphical description.

Using a client-server approach allows you to organize cross-platform user interaction. To access resources on another platform, you only need an appropriate interface agent to interpret the graphics language used by this platform. The client-server principle of interaction allows you to provide both local and remote access to the same resource, thereby increasing the degree of connectivity of the global network.

# Solving tasks in virtual space

## Work technology

Like physical reality, computer virtual reality is an environment for solving user problems. The user manages the resources of this environment using the system management site. Using the tools of this site, tasks are created, the necessary resources are searched, new sites are installed, and the space parameters are changed. A management site consists of a set of workplaces for managing tasks, install local sites, managing devices, changing system settings, and finding environment resources.

The technology of working in a unified environment is basically similar to the technology of a user working in a global network. The basic user model is the widely used browser model. The similarity of spatial structures of the personal domains and data storages provides a similarity of working methods in these structures. The main activities of the user are associated with managing tasks and data in the user's personal domain, as well as solving problems in the space of sites.

The technology of working in a unified environment is basically similar to the technology of a user working in a global network (Figure 10). The main type of user activity is task management, carried out by means of the task management workplace. Quick access to this place can be carried out using the "System" portal, in the form of the "System" key on the keyboard or on the device body.

The technology for solving problems is similar to solving problems in the global network using a browser:
− Create a task portal.
− Search for a site to solve the problem.
− Go to the found site.
− Solving the problem on the site.
− Completion of the task.

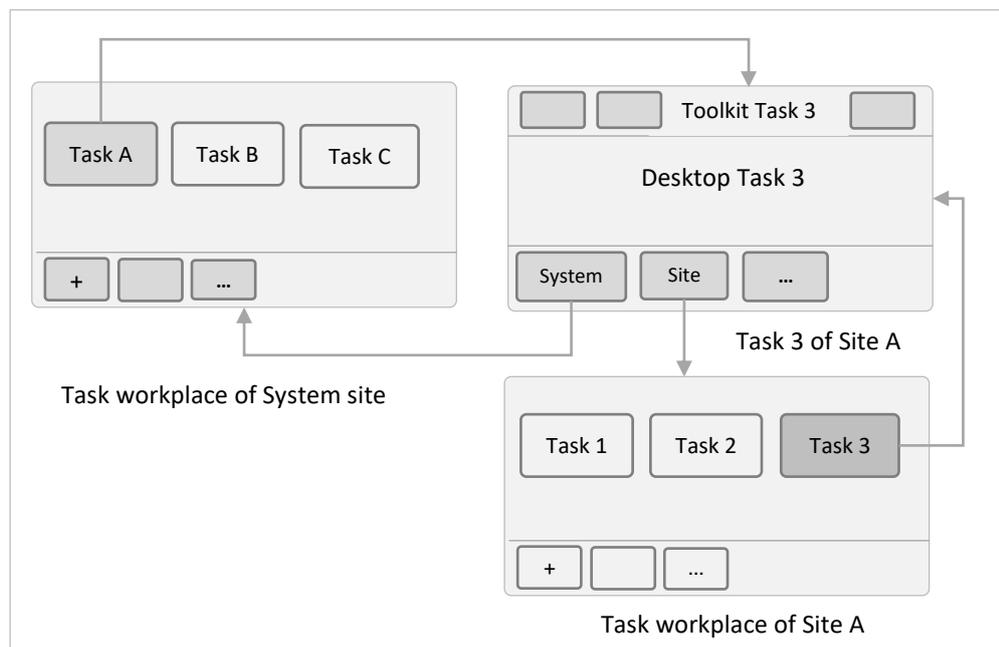

Figure 10. Task control

The task is created using the "Create Task" tool (analogous to the "+" button in the browser). When creating a task, a task portal is created on the task management desktop (analogous to a tab in a browser) and an automatic transition to search workplace with search tools (search engine, environment map, "Favorites" section with links to portals marked by the user, section "History" etc.). Refusal to create a task is possible using the "Undo" tool. Activating the portal of the found site moves the user to the space of that site and associates a task token with that site. To complete the task, you need to apply the "Complete task" tool (analogous to the "×" button in the browser) to the sign of the required task. When multitasking work, switching to the required task is carried out by selecting the portal of the required task on the task table. All actions require no more than two clicks.

Multitasking work on the site is carried out in a similar way, using the tools of the local task management workplace. Quick access to the desktop of this place can be carried out using the "Site" button (local analog of the "System" button), available at any time of work on the site. The task management framework is basically the same as at the system level. Unlike the level of a personal domain, where all tasks are independent, when working on a site, a chain of interrelated subtasks with the transfer of work results is possible. The ability to return from a subtask to the calling task can be carried out using the "Exit" key (similar to the "Escape" tool) with the transfer of the work results (the "Save" tool) or without saving.

Solving problems associated with the use of data objects involves moving objects from storages to the desktops of specialized workstations and vice versa. A new object can be created on the desktop using the "Create Object" tool. You can start working with an existing object by moving it from the storage to the desktop (the "Move object" tool) by finding it using the data management workplace (analogous to the file system Explorer) in the site data workplace. The data storage structure and search methods are similar to the domain space structure and site search methods. You can use parts of other objects for work without changing their structure using the "Move for View" tool. When completing the work, you need to save the results by moving the object from the desktop to the storage (tools "Save in storage" or "Save a version" of the object). Saving can be carried out automatically in the course of work. If problems arise during work, you can restore the results of the work using the command log. To work with objects on the desktop, the tools of the corresponding workplaces are used.

The tasks of changing the structure of the environment are carried out both at a special workplace and in the process of finding the necessary resources. Having found the sign of the desired object and selecting it using the "Select" tool, the user gains access to additional tools for changing the properties of the symbolic object (the "Properties" tool) and moving it in the space of the personal domain ("Delete", "Copy", "Move", "Create Link" ...). You can get a hint about the purpose of any symbolic object using the "What is this?" tool.

### Management tools.

To work in the environment, a set of basic tools with quick access to them is desirable. When working with a desktop PC, they can be installed on the keyboard. When working with tablet PCs, some of them can be installed on the device case, and some in the form of virtual buttons on the monitor screen.

**Task management tools:**
- "System" – for the transition to the task management workstation from solving other system tasks of the personal domain.
- "Site" – for switching to the desktop for managing the tasks of the site.
- "What is this?" – information about the purpose of the object.
- "Find" – search for resources.
- "Select" – selection of an object.
- "UnDo" – cancel the operation
- "ReDo" – to restore the canceled operation.
- "Repeat" – repeats the last performed operation.
- "Save" – save the object.
- "Command" (Function) - command line output.
- "Create task" – create a new task.
- "Complete task" – completion of the task.
- "Exit" (Escape) – return from the subtask to the calling task.
- "Enter" – completes the entry of data or commands.

**Tools for working with a selected object:**
- "Properties" – go to the place to change the properties of the object.
- "Structure" – go to the place of visualization of the structure of the object (View).
- "Move" – move the selected object to the clipboard.
- "Copy" – copy the object to the clipboard.
- "Insert" – insert an object from the clipboard
- "Delete" – move the object to the trash container.

## Conclusion

The transfer of familiar models of the global environment into the structure of local operating system environments will allow unifying the structure of the user's virtual computer environment, turning this environment into a single system built on common principles. The transition to a unified spatial conceptual basis "object - place" will not entail significant changes in the structure of operating systems at the first stage. The proposed solutions require mostly cosmetic changes at the user interface level. Major changes must occur in the mind of the user in connection with the transition to unified mental models and unified terminology. The main task is to completely virtualize the user's environment by separating it from the hardware architecture as much as possible.

The concept of "domain" is becoming a unified way of structuring the user's space. This concept allows not to use the concepts of "computer" and " hardware device" when working. The use of the concept of "domain" in the environment of operating systems integrates the structure of the global and local space in the user's mind. The concept of "personal domain" replaces the concept of "personal computer" in the mind of the user. The concept "partition" in a personal domain replaces concepts "desktop" and external storage devices.

The unified concept "site" as a place for activity and data storage replaces a number of concepts such as "application" (program), "task window", "memory device", and "site" of the global network. In the user's mind, a site is a virtual environment that includes both places for storing data objects and places for working with them. Integration of programs and data within the site is carried out using portals. The concept "program" is replaced by the concept of "toolkit", and the concept of "data" is no longer associated with memory devices. Data becomes virtual objects stored in virtual storage (data sites). The data structure is displayed on special desktops. Special tools are used for data retrieval and logistics. In fact, a data site is another virtual level of data structuring. The introduction of the concept of "storage" allows not to separate the data from the means of managing them. The introduction of the concept "site" into the structure of operating systems environments and the concept of "data site" into the structure of the global network integrates the structure of the global and local space in the user's mind.

The introduction of the concept of "portal" as a means of integrating information necessary for interaction, allows ensuring the methodological homogeneity of the user's work in a unified virtual environment. With this approach, the way of interaction will depend only on the conceptual type of the object and not depend on its location. The portal is also a means of integrating application sites with data sites. With the help of portals, which hide the peculiarities of the interaction process from the user, the user's methods of work in the local environment of his personal domain become similar to the methods of work in the global network.

The similarity of the structures of global domains, personal domains and data sites provides a similarity in the methods of user interaction both with sites and with data at all levels. For the global environment, these are browser methods. For a personal domain environment, these are methods for working with tasks at the level of personal domains and sites. And for storage, these are methods of working with data.

Unification of the virtual environment assumes the use of a single principle of organizing client-server interaction, which provides both local and remote access to all available resources of a unified environment. The use of this approach is desirable, but the transition to the principle of client-server interaction may require a radical change in existing operating systems. In this regard, it is advisable to use this principle in stages. When working with sites on the global network, the basic principle is the client-server principle using HTML interface agents. In an on-premises personal domain environment, both a traditional graphical interface (GUI) and a client-server approach can be used in parallel to facilitate user interaction with local sites. Some sites will continue to interact using the traditional graphical interface, and new sites can be gradually migrated to client-server interaction using the same vector graphics language as the GUI. The user will not distinguish between these methods, since the information about the type of interaction will be hidden by the portals. In existing systems, the parallel use of a graphical interface and HTML clients to organize interaction has long been common practice.